\documentclass[11pt]{article}
\evensidemargin\oddsidemargin
\usepackage{latexsym}

\newcommand{\G}{G\" odel \mbox{}}

\newcommand{\x}{{\bf x}}

{\it }
\begin{document}
\title{\bf  Incompleteness, Complexity, Randomness and Beyond}
\author{Cristian S. \ Calude\\
Department of Computer Science\\
University of Auckland\\
Auckland, New Zealand\\
Email: {\tt cristian@cs.auckland.ac.nz}
}
\date{}
\maketitle

\medskip

\begin{quote}
{\small {\em The Library is composed of an \, \ldots \, infinite number of hexagonal galleries
\, \ldots \, [it] includes all verbal structures, all variations permitted by the twenty-five
 orthographical symbols, but not a single example of absolute nonsense. \, \ldots \, These
phrases, at first glance incoherent, can no doubt be justified in a cryptographical
or allegorical manner; such a justification is verbal and, ex hypothesi, already figures
in the Library. \, \ldots \, The certitude that some shelf in some hexagon held precious
books and that these precious books were inaccessible seemed almost intolerable.
A blasphemous sect suggested that \, \ldots \, all men should juggle letters and symbols until
they constructed, by an improbable gift of chance, these canonical books \, \ldots \,
but the Library is \, \ldots \, useless, incorruptible, secret.}\\
\phantom{xxxxxxxxxxxxxxxxxxxxxxx} Jorge Luis Borges, ``The Library of Babel"
 }
\end{quote}
\medskip

\thispagestyle{empty}

 G\" odel's Incompleteness Theorems have the same scientific status as 
Einstein's principle of relativity, 
Heisenberg's uncertainty principle,    and Watson and Crick's double 
helix model of DNA.  Our aim is to discuss some new faces of the incompleteness phenomenon 
unveiled by an information-theoretic approach to randomness and recent
developments in quantum computing.

\if01
\begin{itemize}
\item Incompleteness and Undecidability
\item Randomness
\item Information-Theoretic Incompleteness
\item Bibliographical Comments
\item References
\item Other Internet Resources
\item Related Entries
\end{itemize}
\fi

\section{Incompleteness and Uncomputability}

Interest in {\bf incompleteness} dates from early times. Incompleteness was an important
 issue for
Aristotle, Kant, Gauss, Kronecker, but it didn't have a fully explicit, 
precise meaning
before the works of Hilbert and Ackermann, Whitehead and Russell,  G\" odel 
and Turing. 

\medskip

In a famous lecture before the {\it International Congress of Mathematicians} (Paris, 1900), 
David Hilbert expressed his conviction of the solvability of every
mathematical problem:
``Wir m\"ussen wissen.  Wir werden wissen."  (We must know. We will
know.).  
Hilbert highlighted the need to clarify the methods of mathematical
reasoning, using a formal system of explicit assumptions, or axioms.
Hilbert's vision was the culmination of 2,000 years of mathematics
going back to Euclidean geometry.  He stipulated that such a formal
axiomatic system should be both `consistent' (free of contradictions)
and `complete' (in that it represents all the truth).

\medskip

 In their
monumental {\it Principia Mathematica} (1925-1927),  Whitehead and Russell developed
the first coherent and precise formal system
aimed to describe the whole of mathematics. Although {\it Principia Mathematica}   held a
great promise  for  Hilbert's demand,  it fell short of actually proving its completeness.

\medskip

After proving the completeness of the system of predicate logic in his doctoral dissertation
(1929),  G\" odel  has
continued the investigation of the completeness problem for more comprehensive formal systems,
especially systems encompassing all known methods of mathematical proof.
In 1931  (see \cite{feferman2})  G\"  odel  proved  his famous  {\it First
Incompleteness Theorem}, which in modern terms reads:
 
\begin{quote} {\it  any computably enumerable, consistent formal axiomatic
system containing elementary arithmetic  is} incomplete, {\it that is, there exist
true, but unprovable (within the system) statements}. 
\end{quote}

\noindent The system is computably enumerable if its `theorems' can be listed by a Turing machine.
Informally, the set of axioms and deduction rules generates all `theorems'; for example, we cannot
take as axioms all true statements about natural numbers as this set is not computably enumerable.
The condition that the system contains the elementary arithmetic is also essential. For example,
the Euclidean geometry which makes statements only about points, circles and lines in general
does not satisfy this condition, hence it might be complete; and, indeed, it is complete as
Tarski has proved. The flat nature of the Euclidean geometry plays no role here,
  non-Euclidean geometries are also complete.
\medskip

This result together with the  Second Incompleteness Theorem (which states that
the consistency of the axioms cannot be proved within the system)
ended a hundred years of attempts to
establish axioms to put mathematics on an axiomatic basis. 
 G\" odel's
 Incompleteness Theorem does not destroy the fundamental idea of formalism, but  suggests that
a) mathematics will be described by many formal systems as opposed to  a universal one, b)
a more sophisticated and  comprehensive  form of formal system than that envisaged by
Hilbert is required  (see also  Post \cite{post}).

\medskip

Anticipating
resistance to his conclusions
\G   wrote   his papers very carefully. 
Speculating on his extreme
caution, Feferman \cite{feferman4} stated that \G ``could have been more centrally involved 
in the development of the fundamental concepts of modern logic -- {\it truth} and {\it
computability} -- than he was."
    \G took pain  to convince various people 
about the validity of his assertions and results, but he avoided any public
debate and considered his results to have been accepted by those whose opinion mattered to
him.  For example, P. Finsler, E. Post and E. Zermelo  were concerned with
priority issues, while  C. Perelman, M.  Barzin, J.  Kuczy\'{n}ski asserted that
\G had in fact discovered another {\it antinomy}; see \cite{dawson}.
Unlike
the others,  Post expressed ``the greatest admiration" for G\" odel's work, conceding that 
``after all
it is not ideas but the execution of ideas that constitute[s]\,\ldots\,  greatness". G\" odel's
result provoked Hilbert's anger,   but he
apparently  accepted its correctness (cf.  \cite{dawson}).   Hilbert never cited  G\"
odel's work.

\medskip

  The reactions of two great philosophers are also of interest. 
Wittgenstein's negative comments (dated 1938 and posthumously  published in ``Remarks on the
foundations of mathematics"  in \cite{bp})  are now  generally considered an embarrassment in the work of a great
philosopher.  Russell realized the importance of G\"odel's
work, but  expressed his continuous puzzlement in a rather ambiguous way
in a letter from 1 April 1963 (addressed to L.  Henkin; see \cite{dawson}): {\it
Are we to think that $2+2$ is not 4, but $4.001$?}. Following the same source, \G remarked (in a letter addressed to A.
Robinson) that ``Russell evidently misinterprets my result; however he does so in a very interesting manner \,\ldots".

\medskip

In the long run  G\" odel's  interpretations of incompleteness prevailed:  the Incompleteness Theorems
neither  rejected the notion of formal system (quite the opposite)  nor  caused despair over
the imposed limitations; they just re-affirmed the  creative power of human reason.
In Post's  celebrated words: ``mathematical proof is [an] essentially creative [activity]."

\medskip

In 1936 Turing \cite{turing}  showed the undecidability of the {\it Halting Problem}, the
question of whether a given computer program will eventually halt:

\begin{quote}
{\it no mechanical procedure  (therefore no  
formal axiomatic theory) can solve the Halting Problem}.
\end{quote}

These two results have very deep connections. To understand them we need to examine 
a very delicate notion: randomness.

\section{Randomness}

What is randomness? Are there random events in nature? Are there laws of randomness?
 Even today, these  few questions  stir controversy.

\begin{quote}
{\em
I am convinced that the
vast majority of my readers, and in fact the vast majority of scientists and even
nonscientists, are convinced that they know what `random' is. A toss of a coin is
random; so is a mutation, and so is the emission of an alpha particle. \,\ldots \, Simple,
isn't it?} said  Kac in \cite{mk}.  
\end{quote}

 Well, no!  Kac knew very well that randomness, the very stuff of life,  could be called many
things, but not simple. The fact that maintaining perfect order 
is difficult surprises no one, but it may come as something of a ``revelation" that
 perfect disorder is 
beyond  reach. People, even
experts, perform poorly when dealing with randomness. The ``gambler fallacy" is
a classical example: the common belief that after a sequence of losses in a game of chance there will
probably  follow
a sequence of gains is false. Various explanations have been suggested: according to one
of them, the human cognitive and psychological constitution, trained over the years to look for patters and trends
(even where there are none) is ``blind" when it comes to see randomness.

\medskip

Randomness is a most troubling
 concept -- it is hard not only
to attain but also to define or even to imagine in spite of the fact that
{\em   have been heroic efforts to
understand randomness} (cf. Efron (cited in Kolata \cite{gina}).

\medskip

Books
on probability theory do not even attempt to define it:  It's
like the concept of a point in geometry books.
According to Beltrami \cite{beltrami}): 

\begin{quote} {\em The subject of probability begins by assuming
that some mechanism of uncertainty is at work giving rise to what is called randomness,
but it is not necessary to distinguish between chance that occurs because of some
hidden order that may exist and chance that is the result of blind lawlessness. This
mechanism, figuratively speaking, churns out a succession of events, each individually
unpredictable, or it conspires to produce an unforeseeable outcome each time a large
ensemble of possibilities is sampled.}
\end{quote}

 Randomness
means the absence of order or  pattern. 
In an extreme sense there is no such notion as
``true randomness". As an illustration note that any sequence (the simplest mathematical infinite object)
 has some kind of 
order, regularity. For example, van der Waerden (1927)  proved that
 {\it in all binary  sequences  at least one of the
two symbols must occur in arithmetical progressions of every length}. Many other patterns common to all sequences have been
subsequently discovered.

\medskip

 Randomness as pattern-breaking (within a given context)
can be viewed in 
 (at least) four ways:

\begin{description}
\item  $\diamond \,$ Randomness as the output of a ``chance" process: 
 patterns are specified by a set of very small
probability. 

\item  $\diamond \,$ Randomness as the result of ``mixing": far-from-equilibrium-states
specify the patterns. 
\item  $\diamond \,$ Randomness as ``mimicking chance": statistical tests specify the patterns.
\item  $\diamond \,$  Randomness as a measure of incompressibility: 
low complexity (short) programs
specify the patterns. 
\end{description}

In what follows, we will focus on the {\it information-theoretic approach to
randomness} proposed by {\it Algorithmic Information Theory}. To this aim we will
work with a fixed alphabet $\Sigma$ and  a universal self-delimiting Turing machine (shortly, universal
Chaitin machine) $ U$ 
processing strings (over  
 $\Sigma$) into strings. Self-delimiting means that no halting program is a prefix of another. 
In this context universality
is a stronger property than classical (Turing) universality: not only can the universal machine
 simulate every other machine, but the simulation is done in the most economical way.
This means that
the  program-size complexity induced by $U$, $H_U(x)$, defined as the length of the shortest program
which on $U$  produces $x$\footnote{Formally, $H_U (x)  = 
 \min \{  |w|  \mid  U(w) = x \}$.}
is asymptotical optimal.\footnote{That is,  for every  Chaitin machine $C,$ there is a constant const
such that for every string $x$ we have $ H_U(x) \le H_C(x) + \mbox{const}.$}

There are various equivalent ways to define the notion of (algorithmic)
random sequence: 
measure-theoretical definitions (Martin-L\" of \cite{martin1,martin} and  Solovay
 \cite{solovaymanu}),   information-theoretical
definitions (Chaitin \cite{chaitin75}
and Schnorr), topological definition  (Hertling and Weihrauch \cite{HW98icalp}).
 For example, an infinite sequence $\x = x_1 x_2 \ldots  x_n
\ldots $   is {\em Chaitin-random} if the difference between the complexity of
a prefix of lenght $n$ and the length itself tends to infinity.\footnote{Formally, $\lim
_{n\rightarrow\infty}H_U(x_1 x_2 \ldots  x_n)-n=\infty$.}

A real $\alpha$ is {\em random }   if its binary
expansion is a random (infinite) sequence
(Chaitin \cite{chaitin75}); the choice of base is
irrelevant (Calude and J\" urgensen \cite{cj}, Hertling and Weihrauch \cite{HW98icalp},
Staiger \cite{staiger}). 

\medskip

Random reals share many properties naturally associated with randomness:

\begin{itemize}
\item a random real has maximum entropy,
\item no random real is  computable,
\item the digits of a random real are `generated' in an unpredictable way,
\item global disorder contrasts with local total order
(any pattern appears).
\end{itemize}

\section{Information-Theoretic Incompleteness}

{\it Is there any relation between randomness and incompleteness}?
The answer is {\it affirmative} and one possibility to reveal such 
relations is to look
at a special class of reals -- the {\it computable enumerable reals}
(see Soare \cite{soare1}). 
 
\medskip

Turing's argument was based on
computable real numbers.  A real
 is {\it computable} if
there is a computable function for calculating its digits
one by one (see Rice \cite{rice}.  There are programs for calculating 
   $\pi$, $e$, $\sqrt{3}$, $\log_2 3$,
all rationals,   all algebraic reals, 
 and in fact all ``natural" constants,  but it is a bit
surprising that {\it nearly all real numbers are not computable}.

\medskip

A real $\alpha$ is  {\em computably enumerable} ({\em c.e.})   if it
is the limit of a computable, increasing, converging
sequence of rationals. 
In contrast with
the case of a computable real, whose digits are given by a computable function,
during the process of approximation of a c.e.  real one may never know how close one is
to the final value. 
Specker \cite{specker}  gave the first example of a {\it convergent, computable sequence of
rationals which does not converge computably}, hence its limit is a {\it c.e. real
which is not computable}.

\medskip

In 1975 a more modern version of the Halting Problem emerged.
Chaitin \cite{chaitin75} introduced the probability that an
arbitrary universal Chaitin machine will eventually halt:

\[\Omega_U=\sum_{U(x) \mbox{  stops}}2^{-|x|}.\]

The number $\Omega_U$ is a probability because  of Kraft's inequality (which applies
to the set of halting programs of the self-delimiting machine $U$).  Chaitin's Omega reals 
 share two apparently
irreconcilable properties: `algorithmic randomness' and `computable
enumerability'.
Note also that c.e. and random reals have many other interesting properties;  for
example, they are  weak truth-table-complete, but not  truth-table-complete (Calude and Nies
\cite{caludenies}).

\medskip

Each $\Omega_U$ depends on the choice of $U$, so there is
not just one Omega (as there is only one $\pi$), but a class of Omegas.
This observation leads to Solovay's question (\cite{solovaymanu}): 
Are there random and computably
enumerable real numbers other than Omegas? 
The answer is {\it negative}, and the
proof is constructive, cf.
Calude, Hertling, Khoussainov, Wang \cite{CHKW98stacs}, Slaman \cite{slaman98} (see also Ku\v{c}era and Slaman
\cite{kucera}):

\begin{quote}
\it  Let $\alpha \in (0,1)$. The
following conditions are equivalent:
\begin{enumerate}
\item[{\rm 1.}] The real $\alpha$ is c.e. and random.
\item[{\rm 2.}] The real $\alpha$ is the halting probability of some universal
Chaitin machine
$U$, 
$\alpha =\Omega_U$.
\end{enumerate}
\end{quote}


To make the discussion more concrete we will formulate all
results relative to $ZFC$, Zermelo-Fraenkel set theory with choice; all theorems 
hold true under more general conditions.
 {\it The First Information-theoretic Incompleteness Theorem}
(Chaitin \cite{chaitin75}) is:

\begin{quote}
 {\it  Let $U$ be a universal Chaitin
machine. Then, $ZFC$, if arithmetically sound,  can prove only finitely many 
statements  of the form $``H_U(x)>m"$. } 
\end{quote}

In fact, {\it there is a constant $c>0$ such that
$ZFC$ cannot prove the statement $``H_U(x)>m"$ if $m > H_U (ZFC) + c$.}
So, all true statements  $``H_U(x)>m"$ (an infinite set)
are unprovable in $ZFC$. Recognizing high complexity is a difficult task even for $ZFC$. The
difficulty depends upon the choice of $U$: 
some $U$'s are worse than others.
Raatikainen \cite{raat} has shown that  {\it there exists a universal Chaitin
machine $U$ so that
$ZFC$, if arithmetically sound, can prove no  statement of the form $``H_U(x)>0"$.}
It follows that $ZFC$, if arithmetically sound, can prove no 
(obviously, true) statement of
the form $``H_U(x)>0"$.

\medskip


 Chaitin's {\it Second 
Information-theoretic Incompleteness Theorem} reads:

\begin{quote}
{\it Let $U$ be a universal Chaitin
machine.
If $ZFC$ is arithmetically sound, then $ZFC$ can determine
the value of only finitely many bits of $\Omega_U$.}
\end{quote}

We can
explicitly compute a bound on the number of bits of $\Omega_U$ which
$ZFC$ can determine, but the bound   {\it is not computable}. For example,
  Chaitin  \cite{ch97} has constructed a  universal Chaitin  
machine $U_{\rm Lisp}$ and a  theory $T$ such  that $T$ can determine the value of
at most $H_{U_{\rm Lisp}} (T) + 15,328$ bits of $\Omega_U$.

\medskip

Can we `find out' the  (finitely many)  bits which $ZFC$ can determine?

\medskip

For every c.e. and random real  $\alpha$ we can construct a  universal Chaitin machine
$U$  such  that  $\alpha = \Omega_U$ and $ZFC$ is able to determine finitely  (but as many as we
want) bits of $\Omega_U$.
Solovay \cite{solovay2k} went into the opposite direction by showing that:

\begin{quote}

{\it We can effectively construct a universal Chaitin machine $U_{Solovay}$ such that $ZFC$, if
arithmetically sound, cannot determine any single bit of  $\Omega_{U_{Solovay}}$.
}
\end{quote}

Chaitin's Second 
Information-theoretic Incompleteness Theorem holds true for any {\it universal} Chaitin machine   while
Solovay constructed a {\it specific} machine.
A  Chaitin machine  for which  Peano Arithmetic can prove its
universality and 
$ZFC$ cannot determine more than the initial
block of 1's of the binary expansion of its halting probability
will be called  {\em Solovay machine}. 
Which c.e. and random reals are halting probabilities
 of Solovay machines?
Calude \cite{crissol} proved the following result:

\begin{quote}
 {\it Assume that  $ZFC$ is arithmetically sound. Then,  every c.e. and random
real is the halting probability of a Solovay machine.}
\end{quote}

For example, if $\alpha \in (3/4, 7/8)$ is c.e. and random, then in the worst case
$ZFC$ can determine its first two bits (11), but no more.
Assume that  $ZFC$ is arithmetically sound. Then, every c.e. and random real 
 $\alpha \in (0, 1/2)$
is the halting probability of a Solovay machine which cannot determine
any single bit of $\alpha$. No c.e. and random real $\alpha \in (1/2, 1)$ has the above
property.

\medskip

A direct consequence of Solovay's result is the following constructive form of
information-theoretic incompleteness:

\begin{quote}
{\it There exists a universal Chaitin machine $U_{Solovay}$ so that
$ZFC$, if arithmetically sound, cannot prove the} true
 statement
 {\rm ``the first bit of $\Omega_{U_{Solovay}}$ is 0".}
\end{quote}

In fact, a more general theorem is true:

\begin{quote}
\it  For every binary string
$s=s_1s_2 \ldots s_n$  we can  effectively construct a Solovay machine
$U_{Solovay}$ such that  the binary expansion of $\Omega_{U_{Solovay}}$ has the string $0s_1s_2
\ldots s_n$ as prefix. 
Hence,  the following  statements
\[{\mbox {\rm ``The }} \,  0^{th}  \,  {\mbox {\rm
 binary digit of the expansion of}}  \, \Omega_{U_{Solovay}} \, {\mbox  {\em is }} \,  0",\]
\[{\mbox {\rm ``The }} \,  1^{st}  \, {\mbox {\rm
 binary digit of the expansion of}}  \, \Omega_{U_{Solovay}} \, {\mbox  {\em is }} \,  s_1",\]
\[{\mbox {\rm ``The }} \,  2^{nd}  \,  {\mbox {\rm
 binary digit of the expansion of}}  \, \Omega_{U_{Solovay}} \, {\mbox  {\em is }} \,  s_2",\]
\[\vdots\]
\[{\mbox {\rm ``The }} \,  n^{th}  \,  {\mbox {\rm
 binary digit of the expansion of}}  \, \Omega_{U_{Solovay}} \,{\mbox  {\em is }} \,  s_n",\]
are true but unprovable in $ZFC$. 
\end{quote}

The information-theoretic version of incompleteness produces, in a constructive way,
 natural examples
in which the axiomatic method is completely powerless. It also shows that 
incompleteness is pervasive, not accidental (for a different approach  see \cite{cjz}).
This may change the general view on the axiomatic method, one of the most
powerful tools in mathematics. In  G\" odel's own words (\cite{godel1}):

\begin{quote}
\em
\ldots besides mathematical intuition there exists another (though only 
probable) criterion of truth of mathematical axioms, namely their 
fruitfulness in mathematics, and one may add, possibly also in physics 
\ldots The simplest case of an application of the criterion  under 
discussion arises when some \ldots axiom has number-theoretical consequences 
verifiable by computation up to any given integer.
\end{quote}

\medskip

Do these results have any impact on mathematics and/or the philosophy of mathematics? 
Opinions vary dramatically.  H. Weyl described incompleteness in a pessimistic way,
as  a {\it constant drain on the enthusiasm} of pursuing scientific research; F. Dyson sees it
in an optimistic way, as an insurance policy that science will go on forever.
And, of course, some would argue that the work of the overwhelming majority
of mathematicians and philosophers has been quite unaffected by the incompleteness
results. One thing is certain: incompleteness has captured the interest of many. Many books
and thousands of technical papers discuss it and its implications
and the  March 29 1999 issue of {\em TIME} magazine  has included \G and Turing  in its list of the
twenty greatest twenty scientists and thinkers of the twentieth century.

\section{Beyond}
In this section we will discuss some recent results which in a way or another ``challenge"
the limits discussed above.

\subsection{Computing a glimpse of an Omega}

Any attempt  to compute the uncomputable or to decide the undecidable is
without doubt challenging,
but hardly new (see, for example,  Marxen and  Buntrock \cite{ marxen},
Stewart \cite{stewart}, Casti \cite{casti97}).  What about computing pieces of
a concrete Omega number?  First, note that any Omega
number is not only
uncomputable, but random, making the computing
task even more demanding.

Computing  lower bounds for Omega is not difficult: we  just  generate
more and more
halting programs. Are the bits produced by such a procedure exact?  {\em
Hardly}.
If the first bit of the approximation happens to be 1, then sure, it is
exact. However, if the
provisional bit given by an approximation is 0, then, due to possible
overflows,
 nothing prevents
the first bit of Omega  to
 be either 0 or 1. This
situation extends to other bits as well. As we have already discussed, only  an initial  run of  1's
may give  exact values
for some  bits of Omega.

Another (more serious) difficulty preventing the computation of a fragment of an Omega
is the following.
{\it Globally}, if we can compute all bits of $\Omega_U$, then we can  solve the
Halting
Problem for every program for $U$ and conversely, knowing all halting programs one
can compute all bits of $\Omega_U$. {\it Locally}, 
given the first $N$ bits of Omega one can decide the halting status of all programs
of length at most $N$. However, if we can solve
for $U$
the Halting
Problem for
all programs up to $N$ bits long we might not  get an exact value for
any bit
of $\Omega_U$ (less all values for the first $N$ bits). Reason: longer
halting programs can
contribute to the value of a ``very
early" bit of the expansion of $\Omega_U$. Using a ``hybrid approach", programming 
combined with mathematical proofs, in \cite{cds} all
 halting programs of up to 84
bits for a concrete $U$  have been calculated. This information has been used to compute  (only)  the
first  63 {\it exact} bits on $\Omega_U$:
\begin{center}
$000000100000010000100000100001110111001100100111100010010011100$.
\end{center}

\subsection{Turing's barrier revisited}
Classically, there are two equivalent ways to look at the mathematical
notion of proof: a) as a finite sequence of sentences strictly obeying some
axioms and inference rules, b) as a specific type of computation. 
Indeed, from a proof given as a sequence
of sentences one can easily construct a machine producing that sequence as the
result of some finite computation and, conversely, giving a machine computing
a proof we can just print all sentences produced during the computation and
arrange them in a sequence.  A proof is an explicit sequence of reasoning steps that
can be inspected at {\it leisure};  {\it  in theory}, if followed with care, such a sequence
either reveals a gap or mistake, or can convince a skeptic of its conclusion, 
in which case the theorem {\it is considered proven}.

This equivalence has stimulated the construction of programs which perform like
{\it artificial mathematicians}.\footnote{Other types of ``reasoning" such as medical diagnosis
or legal inference have been successfully modeled and implemented; see,
for example, the British National Act which has been encoded in first-order logic and a machine
has been used to uncover its potential logical inconsistencies.} From proving
simple theorems of Euclidean geometry to the proof of the four-color theorem, these
``theorem provers" have been very successful. Of course, this was a good
reason for sparking lots of controversies. 
{\it Artificial mathematicians} are far less ingenious and subtle than human mathematicians, but 
they surpass their human counterparts by being infinitely more patient and diligent. 
What about  making errors? Are human mathematicians less prone to errors? This is
a difficult question which requires more attention.

If a conventional proof is replaced by a ``quantum computational
proof" (or a proof produced as a result of a molecular experiment), then the
conversion from a computation to a sequence of sentences may be impossible,
e.g., due to the size of the computation. For example, a quantum machine could be
used to create some proof that relied on quantum interference among all the computations going on
in superposition. The quantum machine would say ``your conjecture is true", but there
will be no way to exhibit all trajectories followed by the quantum machine
in reaching that conclusion. In other words, the quantum machine has the ability
to check a proof, but it may fail to reveal any ``trace" of how it did it. Even worse, any attempt to
{\it watch} the inner working of the quantum machine (e.g. by ``looking"
at  any information concerning the state of the on going proof) may
compromise forever the proof itself!

These facts may not affect the essence of mathematical objects and constructions
(which have an autonomous reality quite independent of the physical reality), but they
seem to have an impact on how we learn/understand mathematics (which is thorough
the physical world). Indeed, our glimpses of mathematics seem to be ``revealed" through physical objects, i.e. human brains, silicon computers, quantum Turing machines, etc.,
hence, according to Deutsch \cite{deutsch-85}, they have to obey not only the axioms and the
 inference  rules of the theory, but the {\it laws of physics} as well. 

The question of trespassing Turing's barrier,
i.e. the possibility to solve a Turing undecidable
problem, to compute an uncomputable function has been considered by various
authors, for example, \cite{hava,jack1,jack2}.  Is there any hope for
quantum (or DNA) computing to challenge the Turing barrier, i.e. to solve an undecidable
problem, to compute an uncomputable function?  According to Feynman's
argument (see \cite{feynman82}, a paper reproduced also in
\cite{feynman99})
any quantum system can be simulated with arbitrary precision by a (probabilistic) Turing machine,  so the answer seems to be {\it negative}.
However, some recent tentative approaches promise a positive answer:  for quantum approaches 
see \cite{cds99,cds00,etesi,cpav,kieu1}\footnote{The  solution proposed in \cite{cpav}
is based on the ``continuity" of quantum programs. Because of
continuity, in deciding
the halting/non-halting status of  a non-halting  machine,  the quantum program
  ``announces" (with a
non-empty probability) the non-halting  decision well before  reaching it;
hence, the
challenge is to design a procedure that detects and measures this  tiny, but
non-empty signal. This was indeed achieved by exploiting some properties of the Brownian  motion. }
and for DNA methods  see \cite{cp}. 

Is incompleteness affected? We need more understanding of the  quantum world to be able to answer this question.
One step toward a possible answer to this question
is to look at the quantum version of  $\Omega$, the number $\Omega_q$ invented
in 1995 by G. Chaitin, K. Svozil and A. Zeilinger (see \cite{karlomega,wc}; see also
\cite{kieu1,vitany}). The number $\Omega_q$  is the probability amplitude with which a random quantum program halts on a  self-delimiting universal quantum
machine  (hence, the halting probability of a self-delimiting universal quantum machine is $|\Omega_q|^2$).\footnote{Things are more complicated as
the halt bit of the quantum machine might enter a superposition state and remain there while other
parts of the output state describing the quantum machine continue to change. Finally, to settle the matter one has to perform a measurement.}  For computing
$\Omega_q$ only the quantum versions of classical bits in the domain of the quantum machine are 
allowed as inputs, so from the computability point of view $\Omega_q$ is an
 $\Omega$,  hence all information-theoretic
results remain unchanged. 
The halting probability of any quantum device capable  to solve the Halting Problem (for classical Turing machines) will
be an $\alpha$ number (as introduced in \cite{veronica}), a
 random, but not c.e. real; the ``incompleteness" derived from such a number has
not (yet) been studied.

As is pointed out in \cite{cds00}, even if theoretically one could show that
Turing's barrier can be trespassed by a quantum machine, the impact on computer technology would be  very low because for all practical purposes the 
halting
computation has a non-zero, but very small chance  of detection.
So, when reality seems so far way from theory, why are we concerned with the later? According to Landauer \cite{land} the answer is:

\begin{quote}
 {\em Because it is at the very core of science. \ldots \, Information, numerical or otherwise, is not an abstraction, but it is inevitable tied to a physical representation. \ldots \,  the handling of information is inevitable tied to the physical universe, its contents and its laws}.
\end{quote}

\section{Digression: Is the Universe Lawless?}

The  hypothesis that the ``Universe is lawful" is supported by our
daily observations: the rhythm of day and night, the pattern of planetary
motion, the regular ticking of clocks.  It is a simple matter of reflection
to point out some limits to this type of argument: the vagaries of weather,
the devastation of earthquakes or the fall of meteorites seem to be fortuitous.  How can the same physical process, for example the spin of
a roulette wheel, obey two contradictory laws, the laws of chance and the
laws of physics?

Perhaps a different hypothesis can better explain this type of behaviour.  As
our direct information refers to {\it finite} experiments, it is not out of
question to discover {\it local rules}, functioning on large, but finite
scales, even if the global behaviour of the process is, or appears to be,
random.  The
fact that the first billion digits of a random sequence are perfectly
lawful, for instance by being exactly the first digits of the decimal
expansion of $\pi$, does not change in any way the global property of
randomness. But, to ``see"  this {\it global randomness} we have to go
beyond the finite; we have to access the  {\it infinite}!  The hypothesis stating 
that the ``Universe is lawless", motivated by a crude
model of  Universe based on the Omega number   developed in
\cite{csal}, was discussed in \cite{cwalt}:   it tries to explain our  {\it partial,
incomplete and provisional}  understanding
of the Universe in a different way. But, of course, the adjectives ``partial,
incomplete and provisional"  apply to the model itself!

\section{Bibliographical Comments}
The list of references is by no means comprehensive and should be used in conjunction
with bibliographies appearing in the cited works.
One of the best presentations of G\" odel's Incompleteness Theorem is \cite{nn}.
The  founders of algorithmic information theory are
Solomonoff \cite{sol1}, Kolmogorov \cite{ko1} and Chaitin \cite{ch1}.
Chaitin's  monographs
\cite{chaitin3,ch98} deal with information-theoretic incompleteness. More on these
issues can be found in \cite{rozarto,barrow,beltrami,zwirn,barrow0,bm}; for critical discussions
see \cite{lam,raat}.
Algorithmic information theory is presented in \cite{ch8,ch97,ch98,ch00,uss,cris,lv}.
For other interesting discussions on randomness see \cite{mk,gina,dem,beltrami,hayes,karl,dwy}.
Easy to understand  presentations include \cite{bg,ch82,casti,cscgjc,crisglimpse,castijim,chow,ch01,rucker}.
Recent literature inspired by G\" odel's incompleteness include \cite{auburn,doxiadis}.
G\" odel's life is discussed in \cite{kleene,kreisel,dawson,wang,castijim}. The literature
on quantum computing is growing at full speed: some references are
\cite{Gruska,wc,cp}.

\section*{Acknowledgment}
We thank Greg Chaitin, Jack Copeland, Fred Kroon, Sergiu Rudeanu, Jerry Seligman and Karl Svozil for useful comments and criticism.

\end{document}